\documentclass[a4paper,12pt]{article}

\usepackage[english]{babel}
\usepackage[utf8x]{inputenc}
\usepackage[T1]{fontenc}
\usepackage{authblk}
\usepackage[dvipdfmx]{graphicx}
\usepackage{natbib}
\usepackage{here}
\usepackage{setspace}
\usepackage{multirow}
\usepackage{lmodern}
\usepackage{amsmath, amssymb}
\usepackage{bm}
\usepackage{pdflscape}
\usepackage{authblk}
\usepackage{color}
\usepackage[table]{xcolor}

\usepackage{booktabs}
\usepackage{afterpage}
\usepackage{textcomp}
\usepackage{float}
\usepackage{enumerate}
\usepackage{rotating}

\definecolor{babyblue}{rgb}{0.54, 0.81, 0.94}
\definecolor{babypink}{rgb}{0.96, 0.76, 0.76}

\usepackage[margin=1in]{geometry}

\usepackage{amsmath}
\usepackage{graphicx}
\usepackage[colorinlistoftodos]{todonotes}
\usepackage[colorlinks=true, allcolors=blue]{hyperref}

\usepackage{caption}
\usepackage{subcaption}

\usepackage{lineno}

\begin{document}

\begin{titlepage}
    \begin{center}
        \vspace*{1cm}
        \large
        \textbf{Assessing soundscape disturbance through hierarchical models and acoustic indices: a case study on a shelterwood logged northern Michigan forest}\\
         \normalsize
           \vspace{5mm}
         Jeffrey W. Doser\textsuperscript{1}, Andrew O. Finley\textsuperscript{1, 2} Eric P. Kasten\textsuperscript{3}, Stuart H. Gage\textsuperscript{4} \\
         \vspace{5mm}
    \end{center}
    \small
         \textsuperscript{1}Department of Forestry, Michigan State University, East Lansing, MI, USA \\
         \textsuperscript{2}Department of Geography, Environment, and Spatial Sciences, Michigan State University, East Lansing, Michigan, USA \\
         \textsuperscript{3}Clinical and Translational Sciences Institute, Biomedical Research Informatics Core, Michigan State University, East Lansing, Michigan, USA \\
         \textsuperscript{4}Department of Entomology, Michigan State University, East Lansing, MI, USA \\
	 \noindent \textbf{Corresponding Author}: Jeffrey W. Doser, telephone: (585) 683-4170; email: doserjef@msu.edu; ORCID ID: 0000-0002-8950-9895

    \section*{Abstract}
    
        Assessing the effects of anthropogenic disturbances on wildlife and natural resources is a necessary conservation task. The soundscape is a critical habitat component for acoustically communicating organisms, but the use of the soundscape as a tool for assessing disturbance impacts has been relatively unexplored until recently. Here we present a broad modeling framework for assessing disturbance impacts on soundscapes, which we apply to quantify the influence of a shelterwood logging on soundscapes in northern Michigan. Our modeling approach can be broadly applied to assess anthropogenic disturbance impacts on soundscapes. The approach accommodates inherent differences in control and treatment sites to improve inference about treatment effects, while also accounting for extraneous variables (e.g., rain) that influence acoustic indices. 
        
        Recordings were obtained at 13 sites before and after a shelterwood logging. Four sites were in the logging region and nine sites served as control recordings outside the logging region. We quantify the soundscapes using common acoustic indices (Normalized Difference Soundscape Index (NDSI), Acoustic Entropy (H), Acoustic Complexity Index (ACI), Acoustic Evenness Index (AEI), Welch Power Spectral Density (PSD)) and build two hierarchical Bayesian models to quantify the changes in the soundscape over the study period. 
        
        Our analysis reveals no long-lasting effects of the shelterwood logging on the soundscape diversity as measured by the NDSI, but analysis of H, AEI, and PSD suggest changes in the evenness of sounds across the frequency spectrum, indicating a potential shift in the avian species communicating in the soundscapes as a result of the logging. Acoustic recordings, in conjunction with this modeling framework, can deliver cost efficient assessment of disturbance impacts on the landscape and underlying biodiversity as represented through the soundscape. 

        \vspace{5mm}

\noindent \textbf{Keywords}: Soundscape, Bayesian, disturbance, ecoacoustics, logging, hierarchical modeling

\end{titlepage}

\section*{Introduction}

Monitoring the effects of natural and anthropogenic disturbances on wildlife and natural resources is an important conservation task \citep{Stem2005}. Assessing the effectiveness of different land management practices is dependent upon the impact of the practices on the surrounding landscape and wildlife using the habitat \citep{franklin1987}. The large-scale growth in remote sensing technology and computationally efficient statistical techniques has greatly improved the  monitoring of vegetation and ecosystem processes across large spatio-temporal regions \citep{finley2019, Zarnetske2019}. However, such remote-sensing methods often do not make general statements about the impact of different anthropogenic activities on habitat and wildlife dynamics. Traditional methods to assess disturbance impacts on wildlife often rely on labor intensive point count surveys and other field collection techniques that often require multiple visits to each site \citep{Royle2004}, are not reliable for cryptic species \citep{Zwart2014}, and cost prohibitive for large spatio-temporal scales. 

Numerous data collection methods (e.g., citizen science \citep{Sun2019}, camera trapping \citep{Karanth1998}) and statistical modeling techniques (e.g., integrated population models \citep{Zipkin2019}, integrated species distribution models \citep{Pacifici2019}) are being explored to deliver cost effective inference about species distributions and population change. Among these alternatives is the burgeoning field of ecoacoustics that provides methods to assess disturbance impacts on wildlife across large spatio-temporal gradients by using long-term soundscape recordings captured via passive acoustic monitoring \citep{gage2017, Burivalova2019}. Soundscapes, defined as the sounds present at a given location at any given point in time \citep{pijanowski:2011, pijanowski-2-2011}, are an important resource for both wildlife and humans. Soundscapes provide a medium for wildlife to communicate to perform essential functions \citep{marler2004, templeton2006, mcgregor2005}, contribute to human well-being and sense of place, and provide inference on landscape interactions and ecological integrity, suggesting the soundscape itself should be managed as a valuable resource \citep{dumyahn2011}. The central idea of soundscape ecology, a sub-domain of the larger field of ecoacoustics, is that the soundscape and its characteristics can reflect changes in ecological status and landscapes, suggesting soundscapes can be used to assess disturbance impacts on wildlife \citep{pijanowski:2011}. Together, soundscape recordings, remote sensing, and computationally efficient statistical methods can provide the means for an efficient environmental monitoring system over large spatio-temporal regions \citep{Pekin2012}. 

Researchers in the field of ecoacoustics have developed numerous acoustic indices to succinctly quantify soundscape recordings \citep{sueur2014, ndsi, villa2011, boelman2007}. Recent work illustrates the potential for soundscape analysis to assess habitat and wildlife disturbance \citep{joo2017, lee2017, deichmann2017, gasc2018, Myers2019, Burivalova2018, Burivalova2019}. \cite{gasc2018} used an acoustic analysis to reveal impacts of a wildfire on wildlife in the Sonoran Desert Sky Islands, while \cite{deichmann2017} revealed impacts of natural gas extraction on biodiversity in tropical forests through the use of soundscape analysis. However, while seen as a potentially useful method for monitoring impacts of forest disturbance on wildlife \citep{Burivalova2019}, there are few examples of its use for assessing impacts of forest management and in particular timber harvesting (i.e., logging). 

Logging has known effects on wildlife communities through the alteration of both the vertical and horizontal habitat structure, among other forest variables \citep{thompson1988, macArthur1958, conner1975, crawford1981, franklin1987}. Avian communities will change according to the specific habitat requirements of each unique species in the community \citep{doyon2005}. Previous approaches to assess such influences have largely been implemented via point-count surveys and other time and labor intensive sampling designs \citep{doyon2005, flaspohler2002, Darras2019}. The use of acoustic recordings to monitor the large-scale changes in wildlife diversity in response to forest management techniques is a cost and time effective alternative. \cite{law2018} recently used passive acoustic recordings to assess impacts of retention forestry on koala populations in Australia. Although use of soundscape analysis may not present species-specific information without manual analysis (although see \cite{Chambert2018}), it can provide long-term and broadscale information regarding the health and diversity of the landscape in response to human activities, such as forest management \citep{pijanowski:2011, pijanowski-2-2011}. Because acoustic monitoring techniques are cost-effective, non-intrusive, and less labor-intensive than more traditional sampling methods, they can serve as a first phase in monitoring programs designed to assess influence of anthropogenic disturbances on wildlife. Given results from acoustic monitoring, researchers and resource management experts can determine if more labor-intensive methods should be used to obtain species-specific information.

Here we provide a novel modeling approach to assess the influence of a shelterwood logging on soundscapes in northern Michigan. Given birds are useful environmental indicators due to their relationship to habitat characteristics \citep{bibby1999, gregory2003}, we focus on soundscapes that occur during the dawn chorus in the month of June, a time when avian species are actively communicating. This modeling approach can be used to provide inference on how disturbance impacts soundscapes themselves, and can also be the first step in assessing the influence of an anthropogenic disturbance on the underlying wildlife.  More specifically, we seek to address the following questions: 

\begin{enumerate}
    \item What is the impact of the shelterwood logging on the soundscapes?
    \item How has the composition of the soundscapes changed as a result of the logging? Are certain frequencies more dominant in soundscapes after the logging than before? 
\end{enumerate}

\section*{Methods}

\subsection*{Study Location and Shelterwood Logging}

\begin{figure}
    \centering
    \includegraphics[width = 15cm]{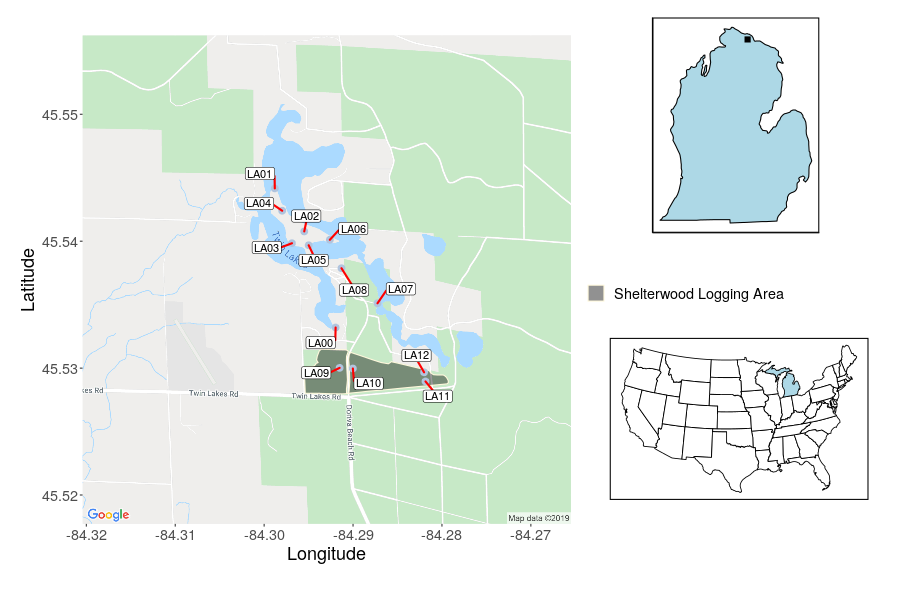}
    \caption{Recording site locations in the Twin Lakes area in Cheboygan, Michigan. Four sites were located in the shelterwood logging area.}
    \label{fig:studyArea}
\end{figure}

Twin Lakes are a chain of seven natural lakes located in the northern lower peninsula of Michigan in Cheboygan county (Figure~\ref{fig:studyArea}). The lakes are primarily groundwater fed, with most basins extending 25-45 ft deep \citep{ndsi}. Most of the land around Twin Lakes is privately owned, but there is a small portion of public land managed by the Michigan Department of Natural Resources (DNR). An important feature of the landscape is an uninhabited island near the center of the lakes (Figure~\ref{fig:studyArea}). Island vegetation consists of 50-60 year old deciduous and coniferous species including white birch (\emph{Betula papyrifera}), trembling aspen (\emph{Populus tremuloides}), balsam fir (\emph{Abies balsamea}), white cedar (\emph{Thuja occidentalis}), tamarack (\emph{Larix laricina}), and white pines (\emph{Pinus strobus}) \citep{Gage2014}. The surrounding forest on public DNR land is natural mixed pine, as well as mixed upland deciduous with conifer. 

The soundscapes of the Twin Lakes area are filled with a variety of acoustically communicating species and anthropogenic activities, and these soundscapes have been the focus of multiple studies from the Remote Environmental Assessment Laboratory (REAL) \citep{ndsi, Gage2014, gage2017}. Previous research on the soundscape temporal variability shows the soundscape does not reflect large amounts of vehicular noise during rush hour periods \citep{Gage2014}. Common amphibians in the area include green frog (\emph{Lithobates Clamitans}), spring peeper (\emph{Pseudacris crucifer}), and northern leopard frog (Lithobates pipiens), while some common avian species include bald eagle (\emph{Haliaeetus leucocephalus}), osprey (\emph{Pandion haliaetus}), Caspian tern \emph{Hydroprogne caspia}), belted kingfisher (\emph{Megaceryle alcyon}), great blue heron (\emph{Ardea herodias}), trumpeter swan (\emph{Cygnus buccinator}), common loon (\emph{Gavia immer}), common merganser (\emph{Mergus merganser}), ruffed grouse (\emph{Bonasa umbellus}), and many smaller woodland birds \citep{ndsi}. Ordinances on the lake prohibit high-speed boating, suggesting the residents and community place value on soundscapes in the area. 

In 2014, a shelterwood logging took place in a well-stocked natural mixed pine forest stand in the Twin Lakes region beginning in late winter 2014 and ending in early spring 2014. Any aspen (\emph{Populus}) and red maple (\emph{Acer rubrum}) present in the area were harvested and red pine was thinned to half-height spacing. The goal of the logging was to obtain regeneration of any combination of aspen, oak (\emph{Quercus}), jack pine (\emph{Pinus banksiana}), red pine (\emph{Pinus resinosa}), or white pine that resulted in a medium-stocked stand. The logging resulted in 73\% basal area removal from the stand \citep{dnr}. Soundscape recordings were subsequently used to assess the influence this logging had on the soundscape.

\subsection*{Soundscape Recording Procedure}

Four sound recorders were placed in the natural mixed pine forest stand exposed to the shelterwood logging one year prior to the harvest (2013) and recorders were placed in the same locations immediately following the logging from 2014-2018. For these four sites, recordings were obtained each year from 2013-2018 using a Song Meter II from Wildlife Acoustics \citep{wildlifeAcoustics}, recording for one minute every 30 minutes (i.e., 48 recordings each day). Recordings were monoaural with a sampling rate of 22,050 Hz and a 16 bit digitization depth using the built-in omnidirectional microphones with a sensitivity of $-36 \pm 4$dB. Recordings were collected throughout the year with the exception of the winter, and were transferred via FTP to the REAL database \citep{ndsi}. For this study, we focus our analysis on recordings taken during the month of June and during the early morning (05:30-07:30). These recordings occur during the dawn chorus during the breeding season (period of highest avian acoustic activity), and thus we can expect the soundscapes to be dominated by the avian community.

Testing whether or not the shelterwood logging has an impact on the soundscape diversity requires comparing soundscapes exposed to the treatment to soundscapes not exposed to the treatment. As ``control'' recordings, we used recordings obtained throughout the Twin Lakes region collected initially for other research purposes. Recordings were obtained at nine sites using the same acoustic recorder and settings as described for the recording sites in the logging region. Six Song Meter SM2 recorders were placed in locations on a remote island in the middle of Twin Lakes. One recorder was placed near residential sites surrounding Twin Lakes, and two recorders were placed on the mainland in undisturbed locations (Figure~\ref{fig:studyArea}). Recordings for these nine sites were taken throughout the years 2009-2018. However, recorder failure and differing study objectives resulted in different numbers of recordings across years and recording locations. Recordings at these nine sites will serve as ``control'' sites, while recordings from the four locations exposed to the shelterwood logging are the ``treatment'' sites, resulting in 13 total recording sites. The number of recordings obtained at each site for each year is shown in Table \ref{tab:recordingNumbers}.

\begin{table}[ht!] 
  \begin{center}
  \caption{Total number of soundscape recordings at each site during each year. Control sites are highlighted in blue and treatment sites are highlighted in pink.}
  \label{tab:recordingNumbers}
  \begin{tabular}{c c c c c c c c c c c c}
    \toprule
    Site & 2009 & 2010 & 2011 & 2012 & 2013 & 2014 & 2015 & 2016 & 2017 & 2018 & Total \\
    \midrule
    \rowcolor{babyblue}
    LA00 & 114 & 150 & 150 & 150 & 150 & 150 & 95 & 150 & 150 & 150 & 1409\\
    \rowcolor{babyblue}
    LA01 & 121 & 150 & 150 & 0 & 150 & 150 & 150 & 150 & 150 & 150 & 1321 \\
    \rowcolor{babyblue}
    LA02 & 140 & 150 & 150 & 0 & 0 & 0 & 0 & 0 & 0 & 0 & 590 \\
    \rowcolor{babyblue}
    LA03 & 150 & 150 & 150 & 150 & 150 & 150 & 150 & 0 & 150 & 590 & 1350 \\
    \rowcolor{babyblue}
    LA04 & 150 & 150 & 150 & 150 & 0 & 0 & 0 & 0 & 0 & 0 & 600 \\
    \rowcolor{babyblue}
    LA05 & 148 & 150 & 150 & 150 & 0 & 0 & 0 & 0 & 0 & 0 & 598 \\
    \rowcolor{babyblue}
    LA06 & 147 & 150 & 150 & 0 & 0 & 0 & 150 & 150 & 120 & 150 & 1018 \\
    \rowcolor{babyblue}
    LA07 & 0 & 0 & 150 & 150 & 150 & 150 & 150 & 0 & 0 & 0 & 750 \\
    \rowcolor{babyblue}
    LA08 & 0 & 0 & 150 & 150 & 150 & 150 & 80 & 0 & 0 & 0 & 680 \\
    \rowcolor{babypink}
    LA09 & 0 & 0 & 0 & 0 & 150 & 110 & 150 & 49 & 130 & 145 & 734 \\
    \rowcolor{babypink}
    LA10 & 0 & 0 & 0 & 0 & 150 & 110 & 150 & 150 & 130 & 145 & 835 \\
    \rowcolor{babypink}
    LA11 & 0 & 0 & 0 & 0 & 150 & 110 & 150 & 150 & 130 & 136 & 826 \\
    \rowcolor{babypink}
    LA12 & 0 & 0 & 0 & 0 & 150 & 110 & 150 & 150 & 140 & 145 & 845 \\
    \bottomrule
    Total & 971 & 1050 & 1350 & 1050 & 1350 & 1190 & 1375 & 1099 & 950 & 1171 & 11556 \\
    \bottomrule
  \end{tabular}
  \end{center}
\end{table}

\subsection*{Soundscape Quantification}

An important question to consider is how to quantify a soundscape and changes in soundscapes in relation to landscape configuration. Acoustic indices are often used as measures of acoustic complexity or diversity, and have been shown to be useful predictors of species richness in temperate terrestrial habitats \citep{Buxton2018, buxton2018a, towsey2014, sueur2014, Rajan2019}.  \cite{fuller2015} assessed the relationship between numerous acoustic indices and levels of landscape fragmentation, and found three acoustic indices, acoustic entropy \citep{Sueur2008}, acoustic evenness \citep{villa2011}, and the Normalized Difference Soundscape Index \citep{ndsi, joo2011}, to correlate well with landscape configuration \citep{fuller2015}. Here, we will use these three indices, as well as the acoustic complexity index \citep{pieretti2011} and 1 kHz binned normalized Welch power spectral densities \citep{welch1967} to quantify the soundscape.

The acoustic entropy index (H) applies the Shannon Diversity index to the temporal and spectral portions of a soundscape recording. Specifically, H is calculated as the product of the temporal and spectral entropy of the acoustic recording, thus incorporating how sound energy is spread throughout the entire recording and throughout the entire frequency spectrum of interest \citep{Sueur2008}. H ranges from 0 to 1, with higher values indicating a higher variety of sounds that are more evenly distributed across frequency bands. 

The acoustic evenness index (AEI) divides a spectrogram into frequency bins and then determines the proportion of sounds above a given threshold in each bin. The Gini index is then applied as a measure of evenness \citep{villa2011}. AEI ranges from 0 to 1, with 0 representing perfect evenness among sounds in the recording and 1 representing a soundscape composed solely of sounds in a single frequency band.

The NDSI computes a ratio between biological and anthropogenic sounds by assuming the sound in the 1-2 kHz band consists of solely anthropogenic noise and the sounds in the 2-11 kHz region consists of only biological sounds \citep{ndsi}. More specifically, the NDSI is computed as 

\begin{center}
    NDSI $= \frac{\beta - \alpha}{\beta + \alpha}$
\end{center}

where $\beta$ is the sum of the 1 kHz binned normalized Welch power spectral density (PSD) \citep{welch1967} from 2-11 kHz (i.e., biological sounds), and $\alpha$ is the normalized PSD of the 1-2 kHz region (i.e., anthropogenic noise). The NDSI ranges from 1 to -1, with 1 representing a diverse soundscape (low anthropogenic noise, high biological sounds) and -1 representing a human-dominated soundscape (high anthropogenic noise, low biological sounds).  

Although the acoustic complexity index (ACI) was not shown to correlate well with changes in landscape \citep{fuller2015}, we incorporate the ACI in our analysis because it has been shown in several studies to correlate well with avian diversity \citep{pieretti2011, fuller2015, Farina2011}, and given the timing of our recordings we expect them to be dominated by avian species. The ACI is calculated using the common observation that biological sounds are more variable than anthropogenic noise. The ACI essentially calculates an average across all frequency bands in a recording of changes in amplitude from one small time step to another \citep{pieretti2011}. The ACI can take any positive value, and takes higher values when the differences between amplitude over small time steps are large, thus giving more weight to variable sounds (i.e., biological sounds) than constant-amplitude sounds (i.e., anthropogenic sounds). 

To further understand how different frequency ranges in the soundscape changed in response to the shelterwood logging, we compute the normalized Welch PSD values for each 1 kHz band from 1-11 kHz. For each soundscape recording, we obtain 10 PSD values that represent the amount of power within each 1 kHz frequency band. We refer to each PSD value as PSD$_i$, which represents the PSD for the $i-(i + 1)$ kHz band.


\subsection*{Model Development}

When working with multiple acoustic indices (or multiple response variables in general), we consider two approaches. First, a univariate model for each index can be developed---delivering important inference for each index. While this may simplify model interpretation and development, it ignores possible correlation among acoustic indices. Thus, a second approach would model all acoustic indices of interest jointly using a multivariate model that accounts for the correlation among the indices. Here we present both a univariate model illustrated using the NDSI and a multivariate model incorporating H, AEI, NDSI, ACI, and the ten PSD variables. 

To determine the effect of a shelterwood logging on soundscapes, we require a method of hypothesis testing that compares the control soundscapes to the treatment soundscapes. The reliability and validity of standard hypothesis testing methods (i.e., ANOVA) are limited in this case because our control sites have differences from the treatment sites (e.g., proximity to water, dominant vegetation, etc.) that likely cause differences in the soundscapes that are not directly related to the shelterwood logging. Thus we require a model that can estimate these differences to improve the reliability in our estimate for the effect of the shelterwood logging, as well as a model that can accommodate the highly unbalanced nature of the data set (Table \ref{tab:recordingNumbers}). To this end, we develop two hierarchical Bayesian models \citep{berliner1996, hooten, clarkTextbook} that accommodate these data characteristics. We first develop a univariate model using the NDSI, and subsequently develop a multivariate model we use with four common acoustic indices (H, AEI, NDSI, and ACI) and ten correlated PSD variables. The hierarchical Bayesian modeling framework provides  benefits, including allowing for direct probability statements regarding parameters of interest as well as yielding statistically robust estimates of uncertainty \citep{hooten, hooten2019, gelman04}. 

\subsubsection*{Univariate Model}

Because NDSI ranges from -1 to 1, we transform the values using a logit transformation to allow for modeling of the NDSI with the normal distribution, which ultimately provides large computational improvements over alternative models (e.g., beta regression \cite{ferrari2004}). The logit transformed NDSI values take the following form: 

\begin{center}
    log($\frac{\text{NDSI} - a}{b - \text{NDSI}})$
\end{center}

where in this case $a = -1$ and $b = 1$. While NDSI can theoretically take values 0 and 1, we have not seen this in practice and thus do not define the logit transformation for these cases. Let $y_{1, i}$ be the logit transformed NDSI value for recording $i$, and let $\bm{y}_1$ be a vector of logit transformed NDSI values for all $N = 11,556$ recordings. $\bm{y}_1$ is ordered by recording site, the day of the recording, the time of the recording, and the year of the recording, respectively. Table~\ref{tab:model1Stacking} provides a visualization of the data ordering used in the subsequent analysis. Exploratory data analysis revealed a complex non-linear trend of $\bm{y}_1$ across time. Although we could model this trend within a state-space model (e.g., \cite{Hines2014}), we instead treat time as a discrete covariate in which we estimate individual year effects for both treatment and control sites.

\begin{table}[ht!] 
  \begin{center}
  \caption{Stacking order of the data vector $\bm{y}_1$. Data are ordered first by site, then day of recording, time of day of recording, then the year of the recording.}
  \label{tab:model1Stacking}
  \begin{tabular}{c c c c c}
    \toprule
    Site & Day & Time & Year \\
    \midrule
    LA00 & 1 & 05:30 & 2009 \\
    LA00 & 1 & 05:30 & 2010 \\
    \vdots & \vdots & \vdots & \vdots \\
    LA00 & 1 & 05:30 & 2018 \\
    LA00 & 1 & 06:00 & 2009 \\ 
    \vdots & \vdots & \vdots & \vdots \\
    LA00 & 1 & 06:00 & 2018 \\ 
    \vdots & \vdots & \vdots & \vdots \\
    LA00 & 2 & 05:30 & 2009 \\
    \vdots & \vdots & \vdots & \vdots \\
    LA00 & 30 & 07:30 & 2018 \\
    LA01 & 1 & 05:30 & 2009 \\
    \vdots & \vdots & \vdots & \vdots \\
    LA12 & 30 & 07:30 & 2018 \\
    \bottomrule
  \end{tabular}
  \end{center}
\end{table}

We now define an ``individual'' soundscape as a soundscape recording at a given location at a given time of day on a specific day (e.g., soundscape at 05:30 at LA01 on June 1 or soundscape at 07:00 at LA12 on June 10, where LA01 and LA12 are unique recording location labels). This individual soundscape is the sample unit with which repeated measures are obtained over different years. We develop a design matrix for each individual soundscape because the number of repeated measures for each soundscape varies. We define $t_i$ as the number of years individual soundscape $i$ was recorded. 

To address potential differences in treatment and control sites, we develop separate design matrices for each type of site (i.e., control vs. treatment). Let $X_{1, i}[, j]$ represent the $j$th column of the design matrix for individual $i$. Using this notation, we define $X_{1, i}[, 1]$ as the overall intercept and define $X_{1, i}[, 2]$ as an indicator variable that takes the value 1 for all treatment sites and 0 for all control sites. Columns $X_{1, i}[, 3] - X_{1, i}[, 7]$ are indicator variables for the year of the specific recording that take value 1 if the recording is in the year represented by a given column and 0 if otherwise. We only include years after the treatment, such that all year effects are interpreted in reference to the values of $\bm{y}_1$ taken prior to the shelterwood logging. This makes the assumption that the soundscapes are stable prior to the shelterwood logging, which was shown to be true for recordings at a subset of these sites in the analysis done in \cite{Gage2014}. $X_{1, i}[, 8] - X_{1, i}[, 12]$ are treatment only year effect indicator variables, such that the value 1 represents a recording at a treatment site in the given year represented by the column, and 0 represents all other scenarios. $X_{1, i}[, 13]$ is an indicator variable for the presence of rain in the recording. The formulation of the design matrices in this manner allows for the following interpretation of the regression coefficients, where $\alpha_{1, j}$ is the regression coefficient corresponding to column $X_{1, i}[, j]$: 

\begin{itemize}
    \item $\alpha_{1, 1}$: intercept across all recordings
    \item $\alpha_{1, 2}$: differences between control and treatment sites that are not a result of the treatment
    \item $\alpha_{1, 3} - \alpha_{1, 7}$: specific year effects (2014-2018), regardless of control or treatment site.  
    \item $\alpha_{1, 8} - \alpha_{1, 12}$: treatment effects for each year after the logging (2014-2018). 
    \item $\alpha_{1, 13}$: effect of rain
\end{itemize}

Thus, we can perform a type of ``Bayesian hypothesis testing'' by determining if the regression coefficients $\alpha_{1, 8} - \alpha_{1, 12}$ are different from 0. By estimating $\alpha_{1, 2}$ we account for inherent differences between the soundscapes of the control sites and the treatment sites, thus increasing our confidence the treatment effects $\alpha_{1, 8} - \alpha_{1, 12}$ represent differences that are the result of the shelterwood logging. 

\begin{figure}
    \centering
    \includegraphics[width = 15cm]{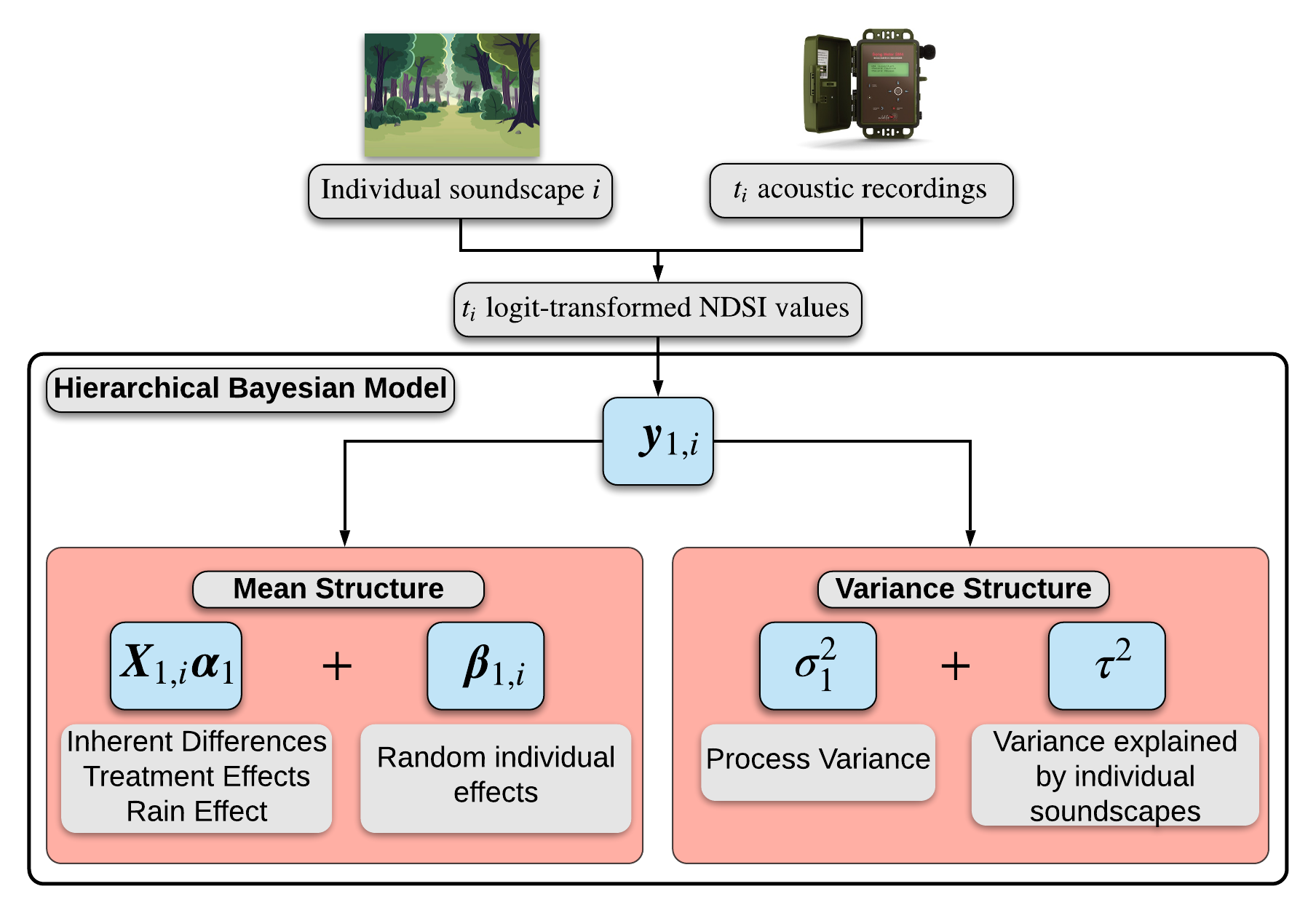}
    \caption{Proposed framework for Model 1. $t_i$ is the number of years individual soundscape $i$ was recorded.}
    \label{fig:univariateModel}
\end{figure}

Bayesian inference focuses on the joint posterior distribution, from which we can make probability statements individually for each parameter on which we seek inference \citep{hooten, hooten2019}. We accomplish this through a hierarchical structure in which we specify a probability distribution for the likelihood, or a data model \citep{hooten}, that takes the form of a basic linear mixed model. Because the data set is unbalanced, we use an individual-by-individual representation of the likelihood through a multivariate normal distribution of dimension $t_i$ for individual $i$. Random effects are implemented for each individual soundscape at the second level of the hierarchy, as we expect differences between these soundscapes as a result of the time of day when the recording occurs, the day of the recording, and the specific site where the sounds are produced. The hierarchical model is completed with the specification of prior distributions for all model parameters to enable estimation within the Bayesian paradigm \citep{gelman04}. We implement an efficient Markov chain Monte Carlo (MCMC) Gibbs sampler as motivated by \citet{ladeau2006} to obtain direct inference on posterior distributions. The model takes the following form: 

\begin{align*}
    [\bm{\alpha}_1, \bm{\beta}_1, \sigma_1^2, \tau^2 \mid \bm{y}_1] & \propto 
	\prod_{i = 1}^{n} \text{Normal}_{t_i}(\bm{X}_{1,i}\bm{\alpha}_1 + \bm{1}_{t_i}\bm{\beta}_{1,i}, \sigma_1^2 \bm{I}_{t_i}) \\
        &\times \text{Normal}_{13} (\bm{\alpha}_1 \mid \bm{0}, 10000\bm{I}_{13}) \\
        &\times \prod_{i = 1}^{n} \text{Normal}(\beta_{1, i} \mid 0, \tau^2) \\
        &\times \text{Inverse Gamma}(\sigma_1^2 \mid 2, 1) \\
        &\times \text{Inverse Gamma}(\tau^2 \mid 2, 1) \\
\end{align*}

where n represents the total number of individual soundscapes, $\bm{\alpha}_1$ are the regression coefficients described previously, $\bm{X}_{1,i}$ is the design matrix for individual soundscape $i$, $\beta_{1, i}$ is the random individual effect for soundscape $i$, $\bm{1}_{t_i}$ is a length $t_i$ vector of ones, $\bm{I}_{a}$ is the $a \times a$ identity matrix, $\sigma_1^2$ is process variance (i.e., variance in the data that is not explained by the design matrix), and $\tau^2$ is random effects variance (i.e., the variance in the data that can be explained by differences in the individual soundscapes). All priors were defined to be non-informative. An overview of the modeling framework is provided in Figure \ref{fig:univariateModel}.

\subsection*{Multivariate Model}

We now develop a multivariate model used to analyze how H, AEI, NDSI, ACI, and the ten normalized PSD values have changed as a result of the shelterwood logging. The model takes essentially the same form as Model 1, except now we estimate all year effects individually for each of the acoustic indices. We model this as a multivariate linear mixed model, in which we have a $14 \times 1$ response vector for each individual recording. Let $\bm{Y}_{2, i}$ denote the $14 \times t_i$ matrix of transformed acoustic indices (log transformation for ACI, logit transformation for all other indices). The design matrix $\bm{X}_{2, i}$ is defined in the same manner as Model 1, except we now estimate all regression coefficients ($\bm{\alpha}_2$) for each index individually, resulting in a total of 182 regression coefficients. Random effects ($\bm{\beta}_2$) are again specified for each individual soundscape, but we estimate all fourteen random effects simultaneously to account for the likely high dependence between the acoustic indices through a covariance matrix $\bm{\lambda}$. This is accomplished using a random effects design matrix $\bm{W}_i$ for each individual. Process error ($\sigma_2^2$) is again estimated as in Model 1. Model estimation comes through a Gibbs algorithm to directly sample from the full conditional distributions for each parameter. Because the model is already multivariate in the sense that we estimate all measures of a given individual soundscape together in the same multivariate normal distribution, we vectorize the response matrix $\bm{Y}_{2, i}$ into the length $14 \times t_i$ vector $\bm{y}_{2, i}$ that contains all acoustic index values for a given individual soundscape for all $t_i$ recording years. The data are stacked in an individual by individual manner analogous to the stacking structure in the univariate model (Table~\ref{tab:model1Stacking}). The structure of the data vector for individual 1 ($\bm{y}_{2, 1}$) is described in Table~\ref{tab:model2Stacking}. All data vectors are formed in an analogous manner for each individual. The full model takes the following form: 

 \begin{align*}
     [\bm{\alpha}_2, \bm{\beta}_2, \sigma_2^2, \bm{\lambda} \mid \bm{Y}_2] &\propto \prod_{i = 1}^{\text{n}} \text{Normal}_{t_i \times 14} (\bm{y}_{2, i} \mid \bm{X}_{2, i}\bm{\alpha}_2 + \bm{W}_i\bm{\beta}_{2, i}, \sigma_2^2\bm{I}_{t_i \times 10}) \\ 
     &\times \text{Normal}_{14}(\bm{\beta}_{2, i} \mid \bm{0}, \bm{\lambda}) \\
     &\times \text{Normal}_{182}(\bm{\alpha}_2 \mid \bm{0}, 10000\bm{I}_{182}) \\
     &\times \text{Inverse Gamma}(\sigma_2^2 \mid 2, 1) \\
     &\times \text{Wishart}(\bm{\lambda}^{-1} \mid (r\bm{R})^{-1}, r)
\end{align*}

where $\bm{R}$ is a 14 x 14 matrix with $0.1$ on the diagonal and 0 elsewhere, and $r$ is the degrees of freedom parameter assigned the value of 14 to create a non-informative prior.

\begin{table}[ht!] 
  \begin{center}
  \caption{Stacking order of the acoustic index values for individual soundscape 1 ($\bm{y}_{2, 1}$). Data are ordered first by site, then day of recording, time of day of recording, year of the recording, then acoustic index.}
  \label{tab:model2Stacking}
  \begin{tabular}{c c c c c}
    \toprule
    Site & Day & Time & Year & Index\\
    \midrule 
    LA00 & 1 & 05:30 & 2009 & H\\
    LA00 & 1 & 05:30 & 2009 & ACI\\
    LA00 & 1 & 05:30 & 2009 & NDSI \\
    LA00 & 1 & 05:30 & 2009 & AEI \\
    LA00 & 1 & 05:30 & 2009 & PSD1 \\ 
    \vdots & \vdots & \vdots & \vdots \\
    LA00 & 1 & 05:30 & 2009 & PSD10 \\
    LA00 & 1 & 05:30 & 2010 & H \\
    \vdots & \vdots & \vdots & \vdots \\
    LA00 & 1 & 05:30 & 2010 & PSD10 \\
    \vdots & \vdots & \vdots & \vdots \\
    LA00 & 1 & 05:30 & 2017 & PSD10 \\
    \bottomrule
  \end{tabular}
  \end{center}
\end{table}

\subsection*{Model Inference and Validation}

Inference from the models primarily focuses on the regression coefficients $\bm{\alpha}$ to provide information on the inherent differences between control and treatment sites ($\alpha_2$), treatment effects ($\alpha_8-\alpha_{12})$, and the effect of rain ($\alpha_{13}$). With regard to the $\alpha$'s, it is common in Bayesian studies to interpret a parameter as ``significant,'' i.e., showing substantial support, if the 95\% credible interval does not include zero, and we take this approach to obtain inference from our models.

Our proposed models include three components that may not be accounted for in more basic hypothesis testing approaches: (1) parameter accounting for inherent differences in the soundscapes; (2) rain effect; (3) random individual effects. As a form of model validation we can look at the values of the parameters estimating these effects and if the 95\% credible intervals do not include zero this provides support for using our modeling approach. Further, we compare how inference from the models changes when removing the three components one at a time from the models, as well as removing all three components at once (referred to as the ``Basic'' model), resulting in four simpler candidate models to compare to our full model. As a more formal method of model selection, we use the widely applicable information criterion (WAIC), a fully Bayesian information criterion \citep{Watanabe2013, hooten}, to compare the five candidate models, with lower values indicating support for that model. Model validation was performed for both the univariate and multivariate model with similar findings, and so we present results solely on the univariate model for the sake of simplicity and brevity. 

\subsection*{Software Implementation}

All data preparation and statistical analysis was performed in R Statistical Software version 3.6.0 \citep{r}. The packages \texttt{soundecology} \citep{soundecology}, \texttt{tuneR} \citep{tuneR}, and \texttt{seewave} \citep{seur2008} were used in computing the acoustic index values, while the \texttt{coda} \citep{coda} package was used in posterior analysis of the model output. All code will be made available upon acceptance or request. Recordings are available through the REAL database \citep{ndsi}.

\section*{Results}

\subsection*{Univariate Model}

\begin{figure}
    \centering
    \includegraphics[width = 15cm]{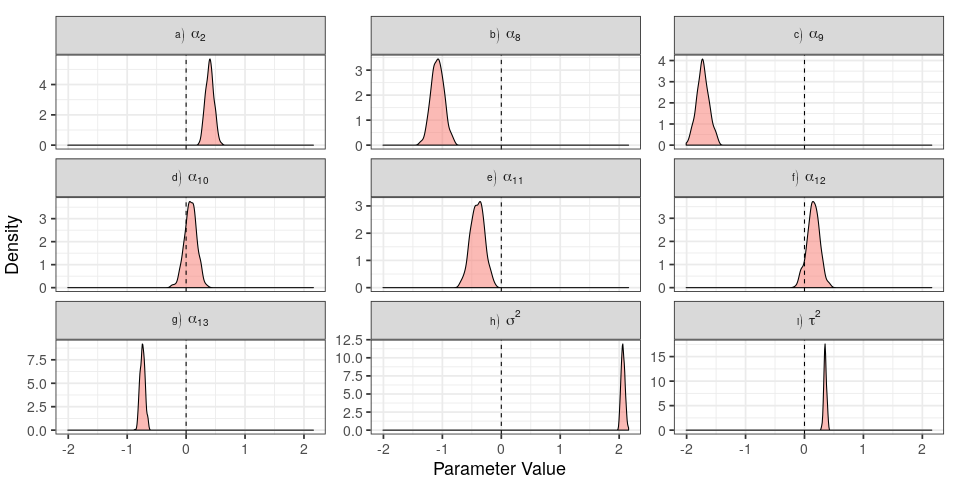}
    \caption{Marginal posterior densities for parameters of interest in the univariate model.}
    \label{fig:univariateGibbs}
\end{figure}

Posterior medians and credible intervals are presented for all parameters in Table \ref{tab:model1Params}. Immediately, we see the process error ($\sigma_1^2$) is much larger than the variability explained by differences in individual soundscapes ($\tau^2$). The process error itself is also very large in comparison to the values of the logit-transformed NDSI, suggesting there is a large amount of variability in the logit-transformed NDSI values that are not accounted for by differences in individual soundscapes, the year effects, or rain. 

The posterior median of $\alpha_{1, 13}$ is less than 0 and the 95\% credible interval does not contain 0, suggesting the presence of rain in the soundscape causes a decrease in the logit-transformed NDSI. In addition, the magnitude of $\alpha_{1, 13}$ is fairly large compared to the estimated year effects, suggesting that rain has an important influence on the values of the logit-transformed NDSI. Rain tends to increase the acoustic PSD in the 1-2 kHz range \citep{Bedoya2017}, and the negative impact of rain on the NDSI (which is inversely related to the 1-2 kHz PSD) aligns with this knowledge.

As described previously $\alpha_{1, 2}$ represents the differences between control and treatment sites that are not a result of the treatment. We see the posterior median of $\alpha_{1, 2}$ is 0.41, and the 95\% credible interval does not contain 0, telling us that there are inherent differences in logit-transformed NDSI at the treatment and control sites. This, however, does not prevent us from drawing inference on the effect of the treatment. In fact, by estimating $\alpha_{1, 2}$, we enhance our confidence in our estimates of the treatment effects because we can be more certain the year effects we estimate are truly a result of the treatment.

\begin{figure}
    \centering
    \includegraphics[width = 10cm]{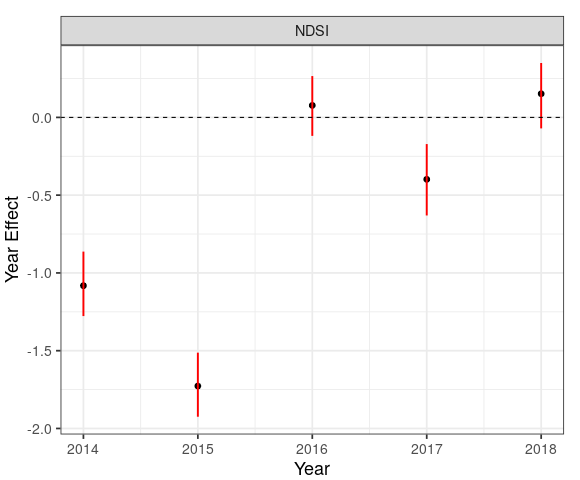}
    \caption{Estimated posterior medians and 95\% credible intervals for treatment effects in each year.}
    \label{fig:ndsiYearEffects}
\end{figure}

Posterior marginal densities of the inherent differences ($\alpha_{1, 2}$), the treatment effects each year ($\alpha_{1, 8}-\alpha_{1, 12}$), the rain effect ($\alpha_{1, 13}$), the random individual effects variance ($\tau^2$), and the process variance $(\sigma_1^2$) are shown in Figure \ref{fig:univariateGibbs}, and the treatment effects for each year are further summarized in Figure \ref{fig:ndsiYearEffects}. We see the effects in 2014 and 2015 are well below zero, suggesting the shelterwood logging caused a decrease in the logit-transformed NDSI values when compared to pre-treatment recordings in 2013. However, in 2016 and 2018, the 95\% credible intervals do overlap zero, and the effect in 2017, although lower than 0, is significantly larger than the effects in 2014 and 2015. This suggests that logging did not have a lasting effect on the soundscape diversity in the Twin Lakes region.

\begin{table}[ht!] 
  \begin{center}
  \caption{Model 1 posterior parameter medians and 95\% credible intervals. Boldface indicates significance (i.e., 95\% credible interval does not contain 0).}
  \label{tab:model1Params}
  \begin{tabular}{c c c c}
    \toprule
    & 50\% & 2.5\% & 97.5\% \\
    \midrule
    $\alpha_{1, 1}$ & \textbf{3.28} & \textbf{3.22} & \textbf{3.33} \\
    $\alpha_{1, 2}$ & \textbf{0.41} & \textbf{0.25} & \textbf{0.52} \\
    $\alpha_{1, 3}$ & \textbf{0.14} & \textbf{0.02} & \textbf{0.24} \\
    $\alpha_{1, 4}$ & \textbf{0.16} & \textbf{0.07} & \textbf{0.27} \\
    $\alpha_{1, 5}$ & \textbf{-0.49} & \textbf{-0.63} & \textbf{-0.36} \\
    $\alpha_{1, 6}$ & \textbf{-0.20} & \textbf{-0.34} & \textbf{-0.04} \\
    $\alpha_{1, 7}$ & \textbf{-0.25} & \textbf{-0.38} & \textbf{-0.14} \\
    $\alpha_{1, 8}$ & \textbf{-1.09} & \textbf{-1.29} & \textbf{-0.89} \\
    $\alpha_{1, 9}$ & \textbf{-1.73} & \textbf{-1.90} & \textbf{-1.52} \\
    $\alpha_{1, 10}$ & 0.08 & -0.11 & 0.32 \\
    $\alpha_{1, 11}$ & \textbf{-0.40} & \textbf{-0.63} & \textbf{-0.18} \\
    $\alpha_{1, 12}$ & 0.15 & -0.06 & 0.35 \\
    $\alpha_{1, 13}$ & \textbf{-0.74} & \textbf{-0.82} & \textbf{-0.66} \\
    $\tau^2$ & \textbf{0.36} & \textbf{0.31} & \textbf{0.40} \\
    $\sigma_1^2$ & \textbf{2.07} & \textbf{2.01} & \textbf{2.12} \\
    \bottomrule
  \end{tabular}
  \end{center}
\end{table}

\begin{figure}
    \centering
    \includegraphics[width = 15cm]{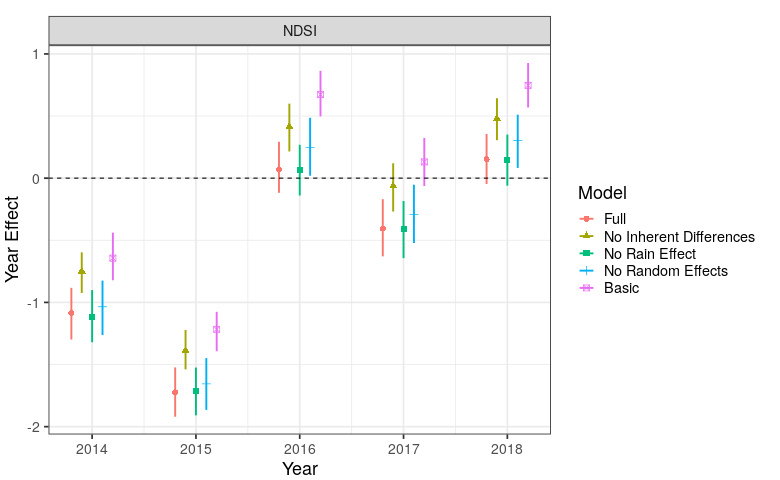}
    \caption{Estimated treatment effects in each year after treatment for each of five models. Points are the posterior median and lines represent the posterior 95\% credible intervals.}
    \label{fig:ndsiValidation}
\end{figure}

Use of our modeling approach is supported by the observation that the 95\% credible intervals for the effect of inherent differences in control and treatment sites ($\alpha_2$) and the rain effect ($\alpha_{13}$) do not contain zero. The random individual effect variance parameter $\tau^2$ is also non-negligible, suggesting it is an important source of variation in the model. Such parameters are often not accounted for in traditional hypothesis testing approaches, which can lead to differing results and interpretation of treatment effects. To display this, we removed these parameters from our model and ran four simpler models: (a) model with no inherent differences parameter; (b) model with no rain effect; (c) model with no random individual effects; (d) model with no inherent differences parameter, no rain effect, and no random individual effects (``Basic'' Model). Results from these models implied different conclusions regarding the effect of the shelterwood logging on the logit-transformed NDSI. Estimated treatment effects from each model are shown in Figure~\ref{fig:ndsiValidation}. Removal of the inherent differences parameter and the random effects results in an increase in the treatment effects in each year. This increase in parameter estimates changes how we interpret the effect of the treatment. Using these more simple models, the decrease in the first two years is not as prominent as in our complete modeling approach. This could potentially have strong implications in management of wildlife and soundscapes if specific values of a decrease in the acoustic index are used as thresholds for disturbance. In addition, the 95\% credible intervals for these two more simple models for years 2016 and 2018 do not contain zero, and so we would interpret the treatment as actually having a \emph{positive} impact on the NDSI. Thus, it is clear the extra complexity incurred in model development by estimating the inherent differences parameter and the random effects parameters is important as not accounting for these parameters can lead to differing conclusions regarding the effect of a treatment. 

From Figure~\ref{fig:ndsiValidation} we see the model that does not include a parameter to estimate the rain effect does not result in any differences in estimated treatment effects, so from a parsimony standpoint including the rain effect is not necessary to draw inferences on the treatment effect. However, estimating this parameter enables us to further understand the influence that rain has on acoustic indices, which is an important task if acoustic recordings and indices are to be used in long-term environmental monitoring programs. Further, we include the estimation of the rain effect parameter in our model to display how other variables that a researcher believes may influence an acoustic index (i.e., wind, planes, temperature) can be included in the modeling framework to provide inference on how such variables impact acoustic indices. In addition, the WAIC was lowest for the full model compared to all other simpler models (Table~\ref{tab:waicValues}), providing further support for the use of our modeling approach. 

\begin{table}[ht!] 
  \begin{center}
  \caption{WAIC values for five candidate models. Lower values indicate better support for a model.}
  \label{tab:waicValues}
  \begin{tabular}{c c c c c }
    \toprule
    Full & No Inherent Differences & No Rain Effect & No Random Effects & Basic \\
    \midrule
    43818.54 & 43895.40 & 44291.36 & 46420.82 & 46977.84 \\
    \bottomrule
  \end{tabular}
  \end{center}
\end{table}

\subsection*{Multivariate Model}

\begin{figure}[ht!]
    \centering
    \includegraphics[width = 10cm, height = 10cm]{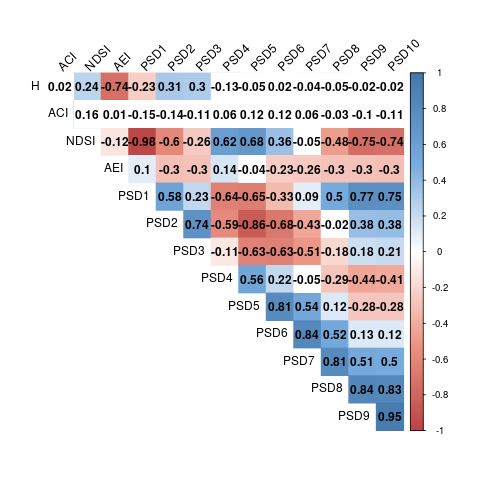}
    \caption{Correlation coefficients for random individual effects of all acoustic indices in the multivariate model. Significance is indicated by non-white color. Red shading indicates significant negative correlation while blue shading indicates significant positive correlation.}
    \label{fig:correlations}
\end{figure}

Correlation coefficients between the random individual effects $\beta_i$ of all acoustic indices are shown in Figure \ref{fig:correlations}. Because the correlation structure is on the random effects and not on the index values themselves, interpretation of the correlation coefficients must be carefully considered. The random effects $\beta_i$ represent the effects of each individual soundscape after accounting for the mean structure of the fixed effects in the design matrix $\bm{X}_i$. This implies high correlation coefficients represent high similarity in the variability of acoustic indices between different individual soundscapes, and thus we can use the correlation coefficients to provide a measure of similarity between the acoustic indices. The 95\% credible intervals for all PSD pairwise combinations, with the exception of three pairs (PSD$_4$-PSD$_7$, PSD$_3$-PSD$_4$, PSD$_2$-PSD$_8$), do not contain zero, suggesting sounds across different frequency ranges are similar to each other. Broadly speaking, the PSD values are positively correlated with PSD values in neighboring frequency ranges, lower frequency PSDs (1-3) are negatively correlated with mid-range PSDs (5-6), and high PSDs (7-10) are highly positively correlated with each other. The NDSI is highly negatively correlated with PSD$_1$, positively correlated with PSD$_4$-PSD$_6$, positively correlated with H, and negatively correlated with AEI. H shows a high negative correlation with AEI, negative correlation with PSD$_1$, and positive correlation with PSD$_2$ and PSD$_3$. AEI is positively correlated with PSD$_1$ and PSD$_4$, and negatively correlated with all other PSD values except PSD$_5$. ACI notably does not show significant correlation with any acoustic index.

The estimated effect of rain on each of the acoustic indices is shown in Table \ref{tab:model2Rain}. Rain has a positive impact on the values of H, no significant impact on ACI, a negative impact on NDSI, and a negative impact on AEI. The rain has an effect on PSD values across the frequency spectrum, with the largest effect being in the 1-2 kHz range, aligning with previous knowledge on the frequencies produced by rain in terrestrial environments \citep{Bedoya2017}.

The parameters estimating the inherent differences between the control and treatment soundscapes that are not the result of the logging are shown in Table~\ref{tab:model2NonLogging}. All indices except for two (ACI and PSD$_6$) had estimates that were different than 0, supporting our claim that the control and treatment sites have inherent differences that need to be accounted for in the modeling framework. 

\begin{table}[ht!] 
  \begin{center}
  \caption{Multivariate model posterior medians and 95\% credible intervals for the rain effect on each transformed acoustic index. Boldface indicates significance (i.e., 95\% credible interval does not contain 0).}
  \label{tab:model2Rain}
  \begin{tabular}{c c c c}
    \toprule
    & 50\% & 2.5\% & 97.5\% \\
    \midrule
    H & \textbf{0.27} & \textbf{0.19} & \textbf{0.35} \\
    ACI & 0.01 & -0.07 & 0.10 \\
    NDSI & \textbf{-0.74} & \textbf{-0.82} & \textbf{-0.66} \\
    AEI & \textbf{-0.43} & \textbf{-0.51} & \textbf{-0.35} \\
    PSD$_1$ & \textbf{0.89} & \textbf{0.80} & \textbf{0.97} \\
    PSD$_2$ & \textbf{0.12} & \textbf{0.03} & \textbf{0.20} \\
    PSD$_3$ & -0.07 & -0.15 & 0.01 \\
    PSD$_4$ & \textbf{0.20} & \textbf{0.12} & \textbf{0.28} \\
    PSD$_5$ & 0.06 & -0.03 & 0.15 \\
    PSD$_6$ & \textbf{-0.12} & \textbf{-0.20} & \textbf{-0.04} \\
    PSD$_7$ & 0.03 & -0.06 & 0.11 \\
    PSD$_8$ & \textbf{0.34} & \textbf{0.25} & \textbf{0.42} \\
    PSD$_9$ & \textbf{0.37} & \textbf{0.28} & \textbf{0.46} \\
    PSD$_{10}$ & \textbf{0.20} & \textbf{0.11} & \textbf{0.29} \\
    \bottomrule
  \end{tabular}
  \end{center}
\end{table}

\begin{table}[ht!] 
  \begin{center}
  \caption{Multivariate model posterior medians and 95\% credible intervals for the inherent differences not a result of the shelterwood logging for each transformed acoustic index.}
  \label{tab:model2NonLogging}
  \begin{tabular}{c c c c}
    \toprule
    & 50\% & 2.5\% & 97.5\% \\
    \midrule
    H & \textbf{0.59} & \textbf{0.47} & \textbf{0.73} \\
    ACI & 0.02 & -0.11 & 0.14 \\ 
    NDSI & \textbf{0.39} & \textbf{0.25} & \textbf{0.54} \\
    AEI & \textbf{-1.72} & \textbf{-1.84} & \textbf{-1.59} \\
    PSD$_1$ & \textbf{-0.31} & \textbf{-0.45} & \textbf{-0.15} \\
    PSD$_2$ & \textbf{-0.64} & \textbf{-0.79} & \textbf{-0.52} \\
    PSD$_3$ & \textbf{0.88} & \textbf{0.75} & \textbf{1.01} \\
    PSD$_4$ & \textbf{0.33} & \textbf{0.22} & \textbf{0.47} \\
    PSD$_5$ & \textbf{-0.39} & \textbf{-0.53} & \textbf{-0.26} \\
    PSD$_6$ & -0.06 & -0.20 & 0.08 \\
    PSD$_7$ & \textbf{0.43} & \textbf{0.29} & \textbf{0.56} \\
    PSD$_8$ & \textbf{0.89} & \textbf{0.75} & \textbf{1.02} \\
    PSD$_9$ & \textbf{0.93} & \textbf{0.80} & \textbf{1.06} \\
    PSD$_{10}$ & \textbf{1.03} & \textbf{0.90} & \textbf{1.16} \\
    \bottomrule
  \end{tabular}
  \end{center}
\end{table}

\begin{figure}
\centering
\includegraphics[width = 15cm]{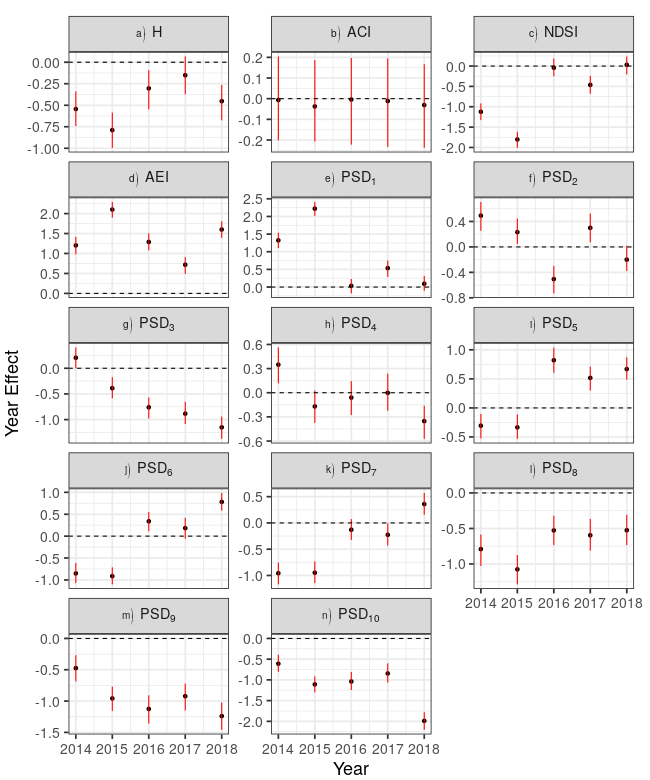}
\caption{Estimated posterior medians and 95\% credible intervals for treatment effects in each year for all transformed acoustic indices in the multivariate model.}
\label{fig:multivariateAlpha}
\end{figure}

Figure \ref{fig:multivariateAlpha} shows the treatment effects for each year on each transformed acoustic index. There are clearly different trends for each acoustic index. H, with the exception of 2017, is lower than pre-treatment values. The ACI remains essentially constant throughout all years. Trends for the NDSI were discussed in the univariate model. AEI is higher than pre-treatment values for all years. The PSD values show clearly different trends, which suggests varying impacts of the shelterwood logging on different frequency ranges of the soundscape. The 1-2 kHz range showed essentially the opposite trend of the NDSI, which makes sense given the inverse relationship between the two indices. Frequencies primarily dominated by the avian community (2-8 kHz) show varying patterns, potentially suggesting a shift in the dominant avian species in the area after the logging. 

\section*{Discussion}

We developed two hierarchical Bayesian models motivated by \citet{ladeau2006} to assess soundscape disturbance from a shelterwood logging in northern Michigan that account for inherent differences in control and treatment sites and readily accommodate both a non-balanced design and missing data. These models can be used with soundscape data as an initial low-cost, low-effort monitoring system for assessing impacts of anthropogenic and natural disturbances on soundscapes. The results from these models can be used to make decisions on whether or not more intensive analysis methods on the soundscape recordings (e.g., manual species recognition, occupancy modeling) or using more traditional techniques (e.g., point-count surveys) are needed to provide data on the disturbance impacts on specific species.

The univariate model revealed the logit-transformed NDSI, a commonly used measure of soundscape diversity in terrestrial ecosystems \citep{fuller2015, Gage2014, gage2017}, was negatively impacted by the shelterwood harvest in the first two years after the logging, but after those first two years the sites within the logging region showed little to no differences compared to the control sites (Figure \ref{fig:ndsiYearEffects}). This suggests the shelterwood logging had an initial influence on the soundscapes, but that these effects were not long-lasting and the soundscape recovered from the initial influence of this logging. The use of acoustics has previously been used in assessments of retention forestry techniques on koala populations \citep{law2018}, however, to our knowledge this is the first study comparing soundscapes exposed to a shelterwood logging with soundscapes not exposed to the logging. Analysis based on the NDSI revealed no long lasting impacts on these specific forest soundscapes, which given the positive relationship between acoustic indices and avian diversity in terrestrial habitats \citep{buxton2018a, Buxton2018} could potentially suggest the logging has no long-term impacts on the species richness in the area, although further research is required to validate this claim. Further research should be performed to explore the influences of other logging techniques on soundscapes in a variety of habitats to provide high level information on the potential impacts different forest management plans have on the animal diversity utilizing the forest soundscapes. With the increasing availability of low cost, programmable acoustic recorders \citep{Hill2019}, this is a cost-effective, non-invasive, and time-effective first step in assessing impacts of forest management techniques on wildlife diversity. 

The univariate model can be used in scenarios when there is only one acoustic index of interest. However, given the high number (> 60) of acoustic indices in published literature \citep{BradferLawrence2019}, it is more reasonable to quantify a soundscape using multiple acoustic indices. The multivariate model we present enables one to obtain inference on multiple acoustic indices at once, while also providing correlation estimates among the individual random effects to provide information on how the acoustic indices are related to each other. We model acoustic entropy (H), the acoustic complexity index (ACI), the Normalized Difference Soundscape Index (NDSI), the Acoustic Entropy Index (AEI), and ten 1kHz wide normalized Welch PSD values in a multivariate model to provide inference on how these measures change as a result of the shelterwood logging. Treatment effects for each index are shown in Figure \ref{fig:multivariateAlpha}. H is lower in soundscapes after the treatment compared to before the treatment, where the first two years show the lowest decrease in H from pre-treatment recordings, and the value in 2017 is not different from pre-treatment recordings. Higher values of H correspond to soundscapes with a larger variety of sounds that are evenly distributed across the frequency bands \citep{sueur2014}. This suggests the soundscape has either a smaller variety of sounds than prior to the logging, or the sounds are less evenly distributed across frequency bands. Analysis of the PSD values reveals the high frequency ranges (PSD8 - PSD10) have shown large decreases after the shelterwood logging, which could have contributed to the decrease in the acoustic entropy. The AEI is also higher for all post-treatment years than prior to the shelterwood logging. Since higher values of AEI correspond to less evenness across frequency bands, we have further support that there is a decrease in the evenness of sounds across frequency bands after the shelterwood logging. The ACI showed no differences across the post-treatment years. This finding corresponds with the results from \cite{fuller2015} who found ACI is not correlated with changes in landscape structure.

Analysis of each 1kHz width PSD value in Model 2 revealed the logit-transformed PSD$_1$ has an an inverse relationship with the NDSI. This suggests the decrease in the logit-transformed NDSI was largely the result of an increase in low-frequency sounds in the soundscape, further supported by the near-perfect negative correlation between random individual effects for the two indices (Figure~\ref{fig:correlations}). The low-frequency portion of a soundscape is typically dominated by anthropogenic noise; however, there are several species that produce sound in this range (e.g., common loon (\textit{Gavia immer})), suggesting the initial increase in PSD$_1$ could be the result of an increased amount of anthropogenic noise or an increase in species communicating in this frequency range. Figure~\ref{fig:multivariateAlpha} also reveals differing trends in the PSD values ranging from 2-8 kHz. Treatment sites appear to have fewer high frequency sounds (8-11 kHz) after the shelterwood logging than prior to the logging. PSD$_5$ and PSD$_6$ show initial decreases followed by values higher than pre-logging values in the final three years of the study. PSD$_3$ shows a steady decline each year after the logging. PSD$_7$ is initially much lower after the logging then shows a steady increase to levels higher than prior to the logging, while PSD$_2$ and PSD$_4$ do not show clear patterns, but appear to be relatively similar to values prior to the logging. These differing trends across the frequency ranges, along with the changes in H and AEI, suggest there are different species using the soundscape within each year, implying the logging caused changes in the specific species in the area. This aligns with previous research on the influences of logging on avian communities that shows different species have unique responses to logging \citep{doyon2005, franklin1987}. 

Figure \ref{fig:correlations} reveals the high correlation between many random individual effects for the acoustic indices used in the multivariate model. We see the ACI is not significantly correlated with any other acoustic index, supporting the claim of \citet{fuller2015} that the ACI does not well reflect changes in landscape configuration. The large correlation between AEI and H random effects  suggests the changes we see in these indices are a result of a decrease in the evenness of sound throughout the frequency spectrum. Together with the changes in the different PSD values, this suggests a potential shift in the avian community as a result of the shelterwood logging, revealing trends that were not evident from the univariate model focusing solely on the NDSI. Different acoustic indices represent different aspects of the acoustic environment, and thus using multiple indices could potentially provide a more complete description of the soundscape \citep{BradferLawrence2019, towsey2014}. In addition, combinations of acoustic indices have been shown to be valuable predictors of acoustically-communicating species richness in temperate terrestrial environments \citep{Buxton2018}. Thus, we recommend the use of the multivariate model to incorporate multiple acoustic indices in future studies of disturbance impacts on soundscape and biodiversity. 

Besides the estimated effects of the logging on the soundscapes across years, our models provide information on the effect of rain on the soundscapes and on the inherent differences in the control and treatment soundscape groups. In these soundscapes, we found rain to have an influence on a majority of the acoustic indices (Table \ref{tab:model2Rain}). By estimating this parameter in our models we account for this large source of variability in the indices. Furthermore, we found the values for the coefficient estimating the inherent differences between the control and treatment soundscapes to be different from 0 in a majority of the acoustic indices (Table \ref{tab:model2NonLogging}). Not accounting for these inherent differences in control and treatment groups could potentially lead to misleading results as the variability explained by the inherent differences would be attributed to the treatment, as displayed for the univariate model in Figure~\ref{fig:ndsiValidation}. Furthermore, our model had the lowest WAIC value compared to four more simple models, suggesting the additional complexities in estimating the inherent differences parameter, rain effect parameter, and the random individual effects improve model fit.

Our analysis provides valuable insights about the influence of a shelterwood logging on soundscapes. Specifically, our analysis reveals no long-lasting effects of the shelterwood logging on the soundscape diversity as measured from the NDSI, but further analysis reveals a decrease in the evenness in sound across different frequency bands, suggesting a potential shift in avian species composition due to the dominance of avian species in the soundscapes. Looking ahead, using soundscape recordings to obtain species specific information could provide valuable insight on whether or not these shifts in frequency correspond to shifts in the avian composition. One potential way of doing this is to extend the model of \cite{Chambert2018} to a dynamic false-positive occupancy model that can provide information on the avian species composition through the use of automated acoustic detection algorithms and small amounts of manual validation. Extending this model would allow for comparison of species occupancy across the different years of the data set. 

Our models can be broadly applied to assess the influence of anthropogenic disturbances (e.g., land development, hydraulic fracturing, oil wells, other forest management techniques, etc.) on soundscapes by comparing a set of control sites with sites exposed to the disturbance over potentially long periods of time. Given the nature of ecological studies, it is often impossible to obtain control sites comparable to the treatment sites. Our models address this challenge by estimating a parameter that represents the differences between the control and treatment sites that are not the result of the disturbance, while also providing inference for an unbalanced data set with large amounts of missing data. By assessing the results from these models, land managers and analysts can determine if further data collection or analysis is required to obtain information on disturbance effects on species of interest. As illustrated in the multivariate analysis, our modeling framework can accommodate a suite of acoustic indices to fully characterize the soundscape. Acoustic data analyzed within the proposed approach can serve as a useful tool for monitoring disturbance impacts on landscapes and biodiversity across both large spatial regions and time periods.

\bibliographystyle{apa}
\bibliography{references}

\end{document}